\documentclass[journal=jacsat,manuscript=article]{achemso}

\usepackage[version=3]{mhchem} 
\usepackage{lmodern}

\author{Xiaohui Tang}
\email{xiaohui.tang@uclouvain.be}
\phone{+32 (0)10472589}
\fax{+32 (0)10472598}
\affiliation[UCL]
{ICTEAM Institute, Universit\'e catholique de Louvain, Place du Levant 3, 1348 Louvain-la-Neuve, Belgium}
\author{Laurent A. Francis}
\affiliation[UCL]
{ICTEAM Institute, Universit\'e catholique de Louvain, Place du Levant 3, 1348 Louvain-la-Neuve, Belgium}
\author{Constantin Augustin Dutu}
\affiliation[UCL]
{ICTEAM Institute, Universit\'e catholique de Louvain, Place du Levant 3, 1348 Louvain-la-Neuve, Belgium}
\author{Nicolas Reckinger}
\affiliation[FUNDP]{PMR, University
of Namur FUNDP, Rue de Bruxelles 61, B-5000 Namur, Belgium}
\author{Jean-Pierre Raskin}
\affiliation[UCL]
{ICTEAM Institute, Universit\'e catholique de Louvain, Place du Levant 3, 1348 Louvain-la-Neuve, Belgium}

\title[\texttt{achemso} demonstration]
{Self-Formation of Sub-10-nm Nanogaps by Silicidation for Resistive Switch in Air}

\begin{document}

\begin{abstract}
{\noindent
We developed a simple and reliable method for the fabrication of sub-10-nm wide nanogaps. The self-formed nanogap is based on the stoichiometric solid-state reaction between metal and Si atoms during silicidation process. The nanogap width is determined by the metal layer thickness. Our proposed method produces nanogaps either symmetric or asymmetric electrodes, as well as, multiple nanogaps within one unique process step for application to complex circuits. Therefore, this method provides high throughput and it is suitable for large-scale production. To demonstrate the feasibility of the proposed fabrication method, nanogap resistive switches have been built and characterized. They exhibit a pronounced hysteresis with up to $10^3$ on/off conductance ratios in air. Our results indicate that the voltages for initially electroforming the device to the switch state are determinated by the nanogap sizes. However, the set and reset voltages of the device do not strongly dependent on the nanogap widths. These phenomena could be helpful to understand how the resistive switching is established.}
\end{abstract}


\newpage
Nanogaps play a very important role in many fields, such as molecular electronics \cite{CK99,XG06}, biological/chemical sensors \cite{HJ10,WX12,AI12}, nanoelectronics \cite{MC10}, nanophotonics \cite{HD11}, nanoscale optoelectronics \cite{CH11}, and many other applications \cite{MH11}. One major challenge in producing nanogaps is the reliable and controllable fabrication of two conducting electrodes with a nanometer-sized separation. Various fabrication methods for nanogaps have been proposed in the literature these last years. They can be classified into two groups according to the electrode nature and properties, hereafter called symmetric and asymmetric electrodes. The former is made of an identical metal while the latter is composed of two different metals. The main fabrication process of symmetric electrodes includes: (i) direct lithography \cite{XL08,YC09,LD10,MM12}, which uses high resolution lithography tools, for example e-beam lithography or focused ion beam etching; (ii) shadow mask evaporation \cite{JL00}, which requires nano objects as masks, such as nanowires, carbon nanotubes and nanofilms; (iii) gap narrowing by electromigration \cite{HP99,DE07}, sputtering \cite{AB97}, electroplating \cite{YK03}, and electrodepositing \cite{JX05} which need additional processes or external circuits; (iv) mechanical breaking \cite{CJ96,JR02}, which is achieved by global lending mechanical force, not promising for manufacturing chips. More recently, some improved methods can be found in the literature, such as nanogap formed by electron-field induction \cite{JH12}, high-current annealing \cite{VS12}, and MeV ion-irradiation \cite{JC12}. These new methods suffer from low yield, high cost, and incompatibility with current complementary metal-oxide-semiconductor (CMOS) process. 
The asymmetric electrodes are interesting and useful in the field of molecular electronics \cite{XC09}. Concerning its fabrication, to our knowledge, only two techniques have been reported. Deshmukh \textit{et al}. \cite{MM03} demonstrated a nanogap between a gold (Au) and a cobalt (Co) electrodes. In their work, a pair of Au electrodes is defined lithographically and then Co is covered onto one of them by solution-phase electrodeposition. The self-limiting feature of the deposition results in the separation of two electrodes by a space of less than 10 nm. The main issue is to prevent the unwanted deposition electrode. Chen \textit{et al}. \cite{XC09} presented a 2-nm-wide nanogap between Au and platinum (Pt) nanorods. They employed an anodic aluminum oxide (AAO) template to chemically synthesize a multisegmented Au-Ni-Pt nanorod. After selectively etching a sacrificial layer of nickel (Ni), which thickness well defines the nanogap width, a nanogap is formed between the Au and the Pt nanorods. The difficult task is to position the Au and the Pt nanorods on the predefined electrodes.

To date, no successful examples have been reported on fabrication of sub-10-nm nanogaps by a controlled process as required for production of very large scale integrated (VLSI) circuits. In this paper, we expose a simple, efficient, and reproducible method for the fabrication of sub-10-nm nanogaps. By the present technique, the nanogap is self-formed in silicidation process without using any high resolution lithography tools and external setups. The nanogap width is well controlled by the metal layer thickness. The process is quite reliable and stable since the formation of the nanogaps is based on the stoichiometric solid-state reaction of metal and Si atoms. Our proposed method provides either symmetric or asymmetric electrodes. Importantly, it can fabricate multiple nanogaps in one unique process for application to complex circuits. Therefore, this method provides high throughput and it is adequate for large-scale production, and is compatible with modern semiconductor technology. In this work, the proposed method is used to fabricate nanogap resistive switches \cite{LC71}, which show impressive memory properties with up to 10$^3$ on/off conductance ratios in air. Our results indicate that the present nanogaps also are of interest for vacuum channel transistor \cite{JW12}, memory and logic devices. 


We limit our analysis of the nanogap formation mechanism to the Pt-silicidation, but other silicide cases can be readily analyzed by making appropriate adaptation. The stoichiometric solid-state reaction of Pt and Si brings two phases, Pt$_2$Si and PtSi, which are governed by different activation energies. The first phase, Pt$_2$Si is unstable and not suitable for electrodes. Figure 1 schematically illustrates the Pt-silicidation procedure in the cross-section view of the electrodes. A Si electrode on insulator substrate is covered by a Pt layer with a thickness $t$, as shown in Fig. 1a. After the silicidation, a PtSi layer forms at the Pt/Si common boundary (line BB'). Several studies have demonstrated that the Pt-silicidation reaction is a two-step process. In the first reaction, Pt atoms in the Pt layer close to the common boundary diffuse into the Si electrode to form Pt$_2$Si. This reaction takes place in 3 seconds when temperature reaches the characteristic temperature at which the transformation from Pt to Pt$_2$Si is fully completed. It is emphasized that Pt atoms far from the common boundary can not participate to the first reaction since the diffusivity of these atoms is not large enough to approach the Si electrode at the characteristic temperature of the silicidation (will be discussed later). In the second reaction, Si diffuses into Pt$_2$Si to form a stable PtSi when the temperature is further increased. During the entire silicidation procedure, a Pt layer of 1 nm consumes a Si layer of 1.32 nm to form a PtSi layer of 1.97 nm \cite{GL03}. As a result, the PtSi thickness is contracted relative to the sum of the Pt and Si thicknesses. As shown in Fig. 1b, the PtSi sidewall surface (line EE') is on the right side of the original line AA', maintaining a nanogap between the Pt layer and the PtSi layer as electrodes. The lack of metal atoms and the contractive PtSi thickness result in a nanogap formation at the common boundary. We assume that during the Pt-silicidation, the dimension change of the PtSi relative to the Si electrode is identical in the three directions and the coverage thickness of the Pt layer for the Si electrode on the sidewall and the top surface is uniform. According to Fig. 1a and 1b, the nanogap width ($d$) equals 0.35 $t$, from $t$ + 1.32 $t$ = $d$ + 1.97 $t$. It is obvious that the nanogap width can be well controlled by the Pt layer thickness. This is the case of asymmetric electrodes, where two electrodes are made from Pt and PtSi, respectively.

On the other hand, the symmetric electrodes of PtSi can be obtained by using the design presented in Fig. 1c. Two Si electrodes with an initial gap $D$' are defined on an insulating substrate. This step can be carried out by current lithography tools since it does not require high resolution. Then a Pt layer with a thickness $t$ is deposited. After the silicidation and removal of the unreacted Pt layer by a selective etching, the gap is narrowed down from the initial width $D$' to the final width $D$ as shown in Fig. 1d. Because the PtSi volume is swelling (expansion) relative to the Si volume, both PtSi sidewalls (lines FF' and EE') are over the original Si sidewalls (lines CC' and BB'), respectively, and approach to the central line AA'. In this case, $D$ = $D$' - 1.3$t$, which is calculated from $D$'+ 2$\times$1.32$t$ = $D$ + 2$\times$1.97$t$. This means that the final width of the nanogap depends on its initial width defined by lithography and also on the Pt layer thickness. 

From the nanoscale structure point of view, the nanogap formation is originated from the molar volume change. In the Pt-silicidation, the stoichiometric solid-state reaction is:
\begin{eqnarray}
\mathrm{Pt} + \mathrm{Si} & \rightarrow & \mathrm{PtSi} \label{eq1}
\end{eqnarray}
For the case of asymmetric electrodes, the volume change $\Delta$${V}$ for the PtSi relative to the elements Si and Pt can be calculated as: 
\begin{eqnarray}
\mathrm{\Delta\textit{V}} & = & \mathrm{\textit{V}_{PtSi}} - (\mathrm{\textit{V}_{Pt}} + \mathrm{\textit{V}_{Si}}) \label{eq2} \end{eqnarray}
By substituting the molar volumes of the PtSi, the Pt element and the Si element into equation (2), where $V$$_\mathrm{PtSi}$ = 18.02 cm$^3$mol$^{-1}$, $V$$_\mathrm{Pt}$ = 9.09 cm$^3$mol$^{-1}$ and $V$$_\mathrm{Si}$ = 12.06 cm$^3$mol$^{-1}$, the molar volume change $\Delta$${V}$ equals -3.13 cm$^3$mol$^{-1}$. This is the reason that the PtSi thickness is contractive relative to the sum of the Pt and Si thickness, while for the case of the symmetric electrodes, it is easy to see that:
\begin{eqnarray}
\mathrm{\textit{V}_{PtSi}} & > & \mathrm{\textit{V}_{Si}} \label{eq3} 
\end{eqnarray}
This is why the nanogap between two Si electrodes is narrowed down when Si is transformed into PtSi through the silicidation process. 

From the atom transfer kinetics point of view, the nanogap formation can be attributed to the Kirkendall effect \cite{KN05,JG10}. Specifically, this is due to the different diffusivities of the Pt and Si atoms moving in and out of the common boundary. At the characteristic temperature, the diffusivity of Pt atoms (2.11 nm$^2$s$^{-1}$) is about 2 orders of magnitude larger than that of Si atoms (7.45$\times$10$^{-2}$ nm$^2$s$^{-1}$) because the activation energy of Pt atom (1.485 eV) is smaller than that of Si atom (1.685 eV) \cite{TS00}. 

The consumed thickness of Pt near the common boundary during the silicidation can be calculated as follows. In the first reaction, the Pt consumption rule is that a Si layer of 0.66 nm consumes a Pt layer of 1 nm to form a Pt$_2$Si layer of 1.43 nm \cite{GL03}. If the formed Pt$_2$Si thickness is known, the consumed Pt thickness can be derived. For example, Fig. 2 points out that at the characteristic temperature of 255$^{\circ}$C, the formed Pt$_2$Si thickness is 32.9 nm, corresponding to a consumed Pt thickness of 23 nm. In other words, Pt atoms 23 nm apart from the common boundary can diffuse into the Si electrode sidewall and participate to the first reaction. The curve in the figure is obtained according to a classical growth rate equation given by
\begin{eqnarray}
\mathrm{\textit{$d$$_{PS}$}^2} & = & \mathrm{\textit{D}_{01}(\textit{dT}/\textit{dt})^{-1}\frac{\textit{k}_{B}}{\textit{E}_{A1}}}\left\{\mathrm{\textit{T}^{2}exp\left(-\frac{\textit{E}_{A1}}{\textit{k}_{B}\textit{T}}\right)} - \mathrm{\textit{T}_0^{2}exp\left(-\frac{\textit{E}_{A1}}{\textit{k}_{B}\textit{T}_{0}}\right)}\right\} 
\label{eq4} 
\end{eqnarray}
Where $d$$_{PS}$ is the formed Pt$_2$Si thickness. $D$$_{01}$ the pre-exponential diffusion coefficient, and $E$$_{A1}$ the average activation energy of Pt atoms. $d{T}/d{t}$ represents the average heating rate, and, $T$ and $T$$_0$ are the characteristics temperature and the starting temperature, respectively. In our process, $D$$_{01}$ = 3 cm$^2$s$^{-1}$, $E$$_{A1}$ = 1.485 eV, $T$ = 255$^\circ$C and $T$$_0$ = 20$^{\circ}$C. $d{T}/d{t}$ =  3.2 Ks$^{-1}$ from RTA at 400$^{\circ}$C for 2 minutes.

To apply the silicidation for self-formed nanogaps, the key condition is contractive silicide thickness. Table 1 summaries the silicide thickness ratio, which is defined by the silicide thickness compared with the sum of metal and Si thicknesses, for some common silicides used in the semiconductor industry. In this table, the metal thickness ($t$$_\mathrm{M}$) is normalized to 1 nm. The consumed silicon thickness ($t$$_\mathrm{Si}$) and the resulting silicide thickness ($t$$_\mathrm{Sil}$) are expressed in nm. The table also lists the positions of the silicide sidewall relative to the original Si sidewall. Although all the silicide phases result in contractive volumes with a ratio lower than one, the reduction of the nanogap width only occurs if the silicide sidewall is over the original Si sidewall. In contrast, if the silicide sidewall is below the original Si sidewall, the nanogap will become wide after the Si electrodes are transformed to silicide electrodes. CoSi$_2$ is classified into this case according to Table 1. 

\textbf{Table 1. Silicide thickness ratios, silicide surface positions relative to original Si surface and main diffusing species during silicide formation.}

\begin{table}[h]
        \centering
        
                \begin{tabular}{l l l l l l l l} 

\hline
\footnotesize{} & \footnotesize{Silicide} & \footnotesize{Metal} & \footnotesize{Consumed} & \footnotesize{Silicide} & \footnotesize{Ratio} & \footnotesize{Silicide} & \footnotesize{Diffusing} \\

\footnotesize{Reference} &\footnotesize{name} & \footnotesize{thickness} & \footnotesize{Si thickness} & \footnotesize{thickness}& \footnotesize{$t$$_\mathrm{Sil}$/} & \footnotesize{position} & \footnotesize{species} \\

\footnotesize{} &\footnotesize{} & \footnotesize{($t$$_\mathrm{M}$)} & \footnotesize{($t$$_\mathrm{Si}$)} & \footnotesize{($t$$_\mathrm{Sil}$)}& \footnotesize{($t$$_\mathrm{M}$+$t$$_\mathrm{Si}$)} & \footnotesize{} & \footnotesize{} \\
\hline

\footnotesize{\cite{XT09}} &\footnotesize{PtSi} & \footnotesize{1} & \footnotesize{1.32} & \footnotesize{1.97} & \footnotesize{0.85} & \footnotesize{Over} & \footnotesize{Pt} \\

\footnotesize{\cite{JC97}} &\footnotesize{TiSi${_2}$} & \footnotesize{1} & \footnotesize{2.22} & \footnotesize{2.44} & \footnotesize{0.78} & \footnotesize{Over} & \footnotesize{Si} \\

\footnotesize{\cite{MObook}} &\footnotesize{CoSi${_2}$} & \footnotesize{1} & \footnotesize{3.61} & \footnotesize{3.98} & \footnotesize{0.76} & \footnotesize{Below} & \footnotesize{Co} \\

\footnotesize{\cite{KN05}} &\footnotesize{NiSi} & \footnotesize{1} & \footnotesize{1.83} & \footnotesize{2.01} & \footnotesize{0.71} & \footnotesize{Over} & \footnotesize{Ni} \\

\footnotesize{\cite{GL07}} &\footnotesize{IrSi} & \footnotesize{1} & \footnotesize{1.42} & \footnotesize{1.99} & \footnotesize{0.82} & \footnotesize{Over} & \footnotesize{Si} \\

\footnotesize{\cite{XT03}} &\footnotesize{ErSi$_{1.7}$} & \footnotesize{1} & \footnotesize{1.09} & \footnotesize{1.67} & \footnotesize{0.80} & \footnotesize{Over} & \footnotesize{Si} \\

\footnotesize{\cite{MObook}} &\footnotesize{YbSi} & \footnotesize{1} & \footnotesize{0.61} & \footnotesize{1.30} & \footnotesize{0.81} & \footnotesize{Over} & \footnotesize{Si} \\

\hline
\end{tabular}
\end{table}  
The other condition is that metal atoms are the dominant diffusing species in silicidation. If Si atoms are the main moving species, which diffuse to the metal, such as the titanium (Ti)-silicidation, as shown in Fig. 3a, the TiSi$_2$ formation is no longer confined to the common boundary. In certain case, the nanogap and multiple bridges between both electrodes are formed together (see Fig. 3b), resulting in short-circuit. The last column of Table 1 lists the dominant diffusing species during the related silicidation. 

In addition, for the asymmetric electrodes, the thickness of the metal layer is very important. Preferably, the Si electrode is fully silicided through carefully calculating the required Pt thickness during the silicidation. If a thinner metal layer is used, an unreacted Si layer will remain on the bottom of the PtSi electrode as shown in Fig. 3c. This is due to the lack of Pt source for the silicidation. For the envisioned application this has no consequence since the PtSi has excellent electrical properties, like low resistivity and high break down current density. However, if a thicker metal layer is used, the nanogap is not formed, because the limited Si source results in the PtSi electrode perfectly covered by a thin Pt layer as shown in Fig. 3d. On other words, the Pt layer is continuous at the common boundary. It is worth noting that the second reaction from Pt$_2$Si to PtSi starts only after all Pt has been consumed. Furthermore, the adhesion of metal and substrate is also very important. For example, in Pt-silicidation, the main issue for the fabrication of the asymmetric electrode is the adhesion of the Pt electrode on the SiO$_2$ substrate. In semiconductor technology, a thin layer of Ti or chromium (Cr) are generally used to promote adhesion. However, Ti and Cr may influence the formation of the PtSi phase in our process. To solve this problem, there are two approaches that can be taken to increase the adhesion. One is to use a self-assemble molecules layer, such as (3-Mercaptopropyl)-trimethoxysilane molecules, as promoter for Pt deposition \cite{DD09}. The second approach is to use oxygen plasma to stimulate or trigger the surface of the SiO$_2$ substrate before the Pt deposition for enhancing the SiO$_2$ surface roughness and facilitating Pt nucleation. 

In summary, the self-formation of nanogaps by silicidation process requires the following conditions: (i) the silicide thickness ratio is less than one, importantly, to obtain a narrower nanogap, the silicide thickness ratio should be close to one; (ii) the silicide sidewall is over the original Si sidewall; (iii) the metal atoms act as the dominant diffusing species during the silicidation; (iv) the thickness of the metal layer is correctly chosen according to its consumption rule; and (v) the final silicide phase has a low resistivity for a good electrical connection. 

In this work, the Pt-silicidation is used to demonstrate the technique, but the same method can be extended to other silicides by appropriately choosing the metal. A commercially available $p$-type silicon-on-insulator (SOI) wafer (resistivity of 15-25 $\Omega$$\cdot$cm) is used as the starting substrate. The thickness of the Si-film is 65 nm on a 145-nm-thick buried SiO$_2$ layer. A Si electrode as shown in Fig. 4a is patterned into the Si-film by optical lithography and reactive ion etching. The narrowest part of the electrode is 12 $\mu$m. A layer of resist is spin-coated on the Si electrode and the buried SiO$_2$ layer. As shown in Fig. 4b, a window including two electrode pads and a wire connecting two electrodes, is opened by optical lithography. The width of the wire is determined by the desired length of the nanogap. To prevent short-circuit between the two electrodes of the nanogap, the wire width is smaller than the narrowest part of the electrode. A 20-nm-thick Pt layer is deposited and the lift-off process is carried out. In Fig. 4b, the Pt wire has a width of 1 $\mu$m. The inset to the figure shows a close view for the common boundary on the Pt wire, where the sidewall of the Si electrode is well covered with the Pt layer, because the deposition angle relative to the substrate surface is well controlled to avoid the shadow effect in this step. This guarantees a continuous Pt wire through the common boundary before the silicidation process. This is also confirmed by the electrical measurements where the resistance between the two electrode pads is found to be ohmic. Finally, the silicidation is processed in a rapid thermal annealing (RTA) system at 400 $^{\circ}$C for 2 minutes under forming gas (N$_2$:H$_2$ 95:5). In this step, when the temperature reaches 255$^{\circ}$C, the first reaction from Pt to Pt$_2$Si is complete. Then, when the temperature is increased up to 338$^{\circ}$C, the second reaction from Pt$_2$Si to PtSi is finished. The silicidation converts Si into PtSi in the regions where Si is covered by Pt. Figure 4c shows the scanning electron microscope (SEM) top-view image close to the common boundary. It is found that the Pt wire is broken at the common boundary and a visible self-formed nanogap is created. The inset to the figure shows the transmission electron microscopy (TEM) cross-section view of the common boundary. The nanogap has an average width of about 8 nm along the 1 $\mu$m-long nanogap, which is perfectly consistent with the calculated value based on the relation $d$ = 0.35$t$, with $t$ = 20 nm. It is seen from Fig. 4c that the nanogap is composed of a heterometallic electrode pair: one electrode in PtSi and the other one in Pt, with the former thicker than the later. It can be found that a thin grey layer remains on the bottom of the PtSi electrode. As mentioned above, it corresponds to the unreacted Si, revealing that the 20-nm-thick Pt film is not sufficient to fully convert 65-nm-thick Si electrode into PtSi. The rough sidewalls observed in Fig. 4d originate from the poor quality of the optical lithography process. However, short-circuit behavior between both electrodes is not found in the measurement with a resistance larger than 10 G$\Omega$, thanks to the stoichiometric solid-state reaction of Pt and Si atoms. The overall shape of the Pt electrode sidewall is completely mirrored by the PtSi electrode sidewall. The convex parts on the Pt electrode sidewall create the concave parts on the PtSi electrode sidewall, while maintaining the nanogap along the length of the electrode sidewall. This avoids extremely small nanogaps being filled by defects and irregularities, thereby avoiding short-circuit. This is one great advantage of this process. 

Up to now, the multiple nanogaps are formed by individually undergoing. Although Johnston \textit{et al.} \cite{DE07} used an electromigration technique to fabricate the multiple nanogaps in parallel, a special electrode design is required to ensure the junction resistance larger than the interjunction resistance. As shown in Fig. 4e, our proposed method can fabricate large numbers of nanogaps within one unique silicidation process step so that it is suitable to large-scale production and complex circuit application. This is another clear advantage of this process. We also test the Ti-silicidation to verify the importance of the metal diffusing species. Figure 4f shows a close-view of the SEM image for the common boundary between Ti and TiSi$_2$ electrodes. During TiSi$_{2}$ formation, Si atoms, as the main moving species, can diffuse a long distance and form TiSi$_2$ in the nanogap to make short-circuit bridges between Ti and TiSi$_2$ electrodes. This result confirms the fact that the choice of the metal is very important in the self-formation of nanogap by silicidation. 

The other objective of the present work is to fabricate symmetric electrode nanogaps. All the fabrication steps are the same as that of the asymmetric electrode nanogaps with the following exceptions. Two Si electrodes separated by a gap are defined on the buried SiO$_2$ substrate. The unreacted Pt is removed after the silicidation by an aqua regia solution at room temperature. Figure 5a and 5b show two samples of patterned gaps between a Si electrode pair. Their initial width ($D$'), from the apex point of one electrode to the apex point of the other one, is 60 nm. A Pt layer with a thickness of respectively, 25 nm and 30 nm is deposited on top of these Si electrodes. After the silicidation and removal of the unreacted Pt layer, the final widths ($D$) of both gaps are 14 nm and 6 nm, respectively, as shown in Fig 5c and 5d. Based on the relation $D$ = $D$' - $\sqrt{2}$$\times$1.3$\times$$t$, where $t$ = 25 nm and 30 nm, we can calculate $D$ to be 14 and 6 nm, respectively. These calculated results are in good agreement with the measurement values. It is worth noting that the expansion of the Si electrode along to $x$ and $y$ directions in the wafer plane is considered by a multiplicative factor of $\sqrt{2}$. Our results confirm that the nanogap can be narrowed down after the Pt-silicidation. Importantly, the reduction of the nanogap width can be well controlled by the thickness of the deposed Pt layer. 
 
Pt and PtSi are well known and accepted in the advanced CMOS process, such as in the fabrication of Schottky-barrier metal-oxide-semiconductor field-effect transistors (MOSFETs) \cite{XT08}. Today's high density non-volatile memories are facing tough barriers for scaling down to 10 nm. The resistive switch is believed out as one of the most competitive candidates to replace these memories due to its simple structure (two terminals), high integration density (4F$^2$, F is the minimum feature size) \cite{SH11}, high speed (300 picoseconds) \cite{TH11}, excellent endurance ($>$10$^{12}$ cycles) \cite{CB11}, reliable retention ($>$10 years) \cite{SY11} and compatibility with CMOS process. In addition, Pt has a good surface chemical stability so that it is frequently used as the electrode material of resistive switches \cite{HS12}. In this work, we directly used the fabricated nanogaps with symmetric electrodes as lateral resistive switches, it means without including any switching materials. To test their performance, two PtSi squares (as shown in Fig 5c and 5d) serve as electrodes of the resistive switch. Both electrodes are separated by an air nanogap. In the resistive switch, the set means the transition from high resistance state (HRS) to low resistance state (LRS), while the reset represents the transition from LRS to HRS. The diagonals of two squares are aligned to reduce the power consumption of the resistive switch since the set and reset voltages depend on the area of facing section between the electrodes \cite{JZ10,HM11}. The top surface area of each square is 100 x 100 $\mu$m$^2$ for an easy probing. Electrical characterization is performed by HP4156C semiconductor parameter analyzer at room temperature in air.
 
Figure 6a demonstrates typical current-voltage curves of a nanogap resistive switch with current compliance at 500 nA, measured in air. The nanogap has a width of 10 nm. The black rectangles and red circles show the results obtained when the voltage is changed from 0 to 10 V and from 10 to 0 V, respectively. A pronounced hysteresis is observed and the threshold voltage is defined by the point where a sharp current jump appears (V$_{th}$ in the figure). This indicates that the set and reset processes are achieved in this nanogap. The switching ratio of the device is greater than 10$^3$. Comparing our nanogap resistive switch with the literature \cite{HS12}, the reset process is remarkable. This may be associated with the property and shape of the PtSi electrodes. In our devices, the set$/$reset voltages do not have a strong dependence on the nanogap sizes which is similar to the phenomena reported by Yang \textit{et al}. \cite{JJ09}. 

It is must be underlined that an initial electroforming is necessary to activate the pristine device to a switching state \cite{JY10}. Figure 6b and 6c show the electrical characterization in the initial formation process for two resistive switches with a nanogap width of 10 nm and 14 nm, respectively. This process is performed by continuously sweeping voltage from 0 to 30 V and from 30 to 0 V with a certain current compliance. Here, the voltage stress is used to induce a soft breakdown of the buried SiO$_2$ layer underneath the electrodes for the filament formation \cite{MB11} /the electrochemical reaction \cite{KN11} or to adjust the Schottky barrier between the PtSi electrodes around the nanogap and the buried SiO$_2$ layer \cite{QX11}. The real mechanism of the resistive switch is still under investigation and further experiment is necessary. Conveniently, our resistive switch is a lateral nanogap, which is a promising structure to easily investigate the nanoscale physical and chemical origins of the resistive switch. Comparing the current-voltage curves of two resistive switches, the off-to-on and the on-to-off transition voltages are significantly decreased for the resistive switch with the narrower nanogap. It can also be seen that the narrower the nanogap, the smaller the window between $V$$_\mathrm{ON}$ and $V$$_\mathrm{OFF}$ is. These phenomena could be helpful to understand how the resistive switching is established. Finally, our results show that silicides, such as PtSi, can be of use as electrode materials for resistive switches.

In conclusion, nanogaps with a width of a few nanometers have received considerable attention recently in the scientific and engineering communities. This work presents a simple method to obtain sub-10-nm wide nanogaps without using any expensive lithography tools. The nanogaps are self-formed during the silicidation process. This method shows distinctive advantages: the nanogap width is precisely determined by the metal thickness; the stoichiometric solid-state reaction of metal and Si atoms ensures the reliability and reproducibility of the fabrication; the process is based on the excellent control of the rapid thermal annealing system and large numbers of nanogaps can be fabricated in one unique silicidation process step. The process methodology can be easily adapted to other silicides by appropriately choosing the metal. This method can produce multiple nanogaps either symmetric or asymmetric electrodes for complex circuits with high yield rate (100\%). This work shows that the fabricated nanogaps can be directly used as a lateral resistive switch working in air, which exhibits impressive memory properties with up to 10$^3$ on/off conductance ratios. Our results indicate that the voltages for initially electroforming the device to the switch state are determined by the nanogap size. However, the set and reset voltages of the device do not have a strong dependence on the nanogap width. These phenomena could be helpful to understand how the resistive switching is established. Finally, our results show that silicides, such as PtSi, can be of use as electrode material for resistive switches.

\section{Acknowledgement}
We thank all the engineers and technicians in the UCL-WINFAB for technical support. Xiaohui Tang is a senior researcher of the F.R.S.-F.N.R.S.

\newpage

\newpage
{Figure 1:} {Schematic cross-sectional views for the formation mechanism of nanogaps: (a) a Si electrode covered by a Pt layer on an insulator substrate before the Pt-silicidation; (b) an asymmetric nanogap is self-formed based on the solid-state stoichiometric reaction between Si and Pt atoms during the Pt-silicidation; (c) two Si electrodes with a gap covered by a Pt layer; (d) after Pt-silicidation, the gap narrows down due to the larger PtSi molar volume relative to the Si molar volume.} \\

{Figure 2:} {The formed Pt$_2$Si thickness as a function of temperature in the first reaction of the Pt-silicidation. The calculation is performed by using \textit{D}$_{01}$ = 3 cm$^2$s$^{-1}$, \textit{E}$_{A1}$ = 1.485 eV, \textit{dT/dt} = 3.2 Ks$^{-1}$ and \textit{T}$_0$ = 20 $^\circ$C}. \\

{Figure 3:} {Schematic cross-sectional views for the formation conditions of nanogaps: (a) a Si electrode covered by a Ti layer on an insulator substrate before the Ti-silicidation; (b) During the Ti-silicidation, Si atoms diffuse into the Ti layer to form TiSi$_2$ multiple bridges in the nanogap; (c) a Si electrode covered by a thinner Pt layer, after the Pt-silicidation, an unreacted Si layer remains on the bottom of the PtSi electrode; (d) a Si electrode covered by a thicker Pt layer, after Pt-silicidation, the top and sidewall of the PtSi electrode is covered by a thin Pt layer, no nanogap is found.} \\

{Figure 4:} {SEM top-views and TEM cross-sectional views for the formation of asymmetric nanogaps: (a) a Si electrode defined on a SiO$_2$ substrate; (b) a Pt layer, including a wire connecting the two electrodes, deposited on the Si electrode and the substrate. The inset zooms in on the common boundary of the Pt wire which is continuous; (c) a nanogap is self formed between the Pt electrode and PtSi electrodes. The inset is TEM cross-section of the nanogap; (d) the overall shape of the Pt electrode sidewall is mirrored by the PtSi electrode sidewall, preventing bridge formation between both electrodes; (e) a cross-sectional TEM picture for the multiple nanogaps fabricated in one unique Pt-silicidation process; (f) Mutiple bridges are formed in the nanogap between TiSi$_2$ and Ti electrodes, resulting in short-circuit.} \\

{Figure 5:} {Top views of the SEM images for the formation of symmetric nanogaps and the dependence of the nanogap width on the Pt layer thickness: (a) and (b) two Si electrode pairs patterned with an original gap width of 60 nm. (c) and (d) a Pt layer with a thickness of 25 nm and 30 nm is deposited on the electrodes in (a) and (b), respectively. The nanogap widths are reduced to 14 nm and 6 nm, respectively, after the silicidation.} \\

{Figure 6:} {Typical current-voltage curves with current compliances for a nanogap resistive switch measured in air: (a) a pronounced current hysteresis, corresponding to set and reset states with a threshold voltage of about 8.5 V; (b) and (c) electroforming to activate a switching state for resistive switches with a nanogap width of 10 nm and 14 nm, respectively.} \\

\newpage

\providecommand*\mcitethebibliography{\thebibliography}
\csname @ifundefined\endcsname{endmcitethebibliography}
  {\let\endmcitethebibliography\endthebibliography}{}

\newpage
\begin{figure}
\begin{center}
\includegraphics[width=16cm]{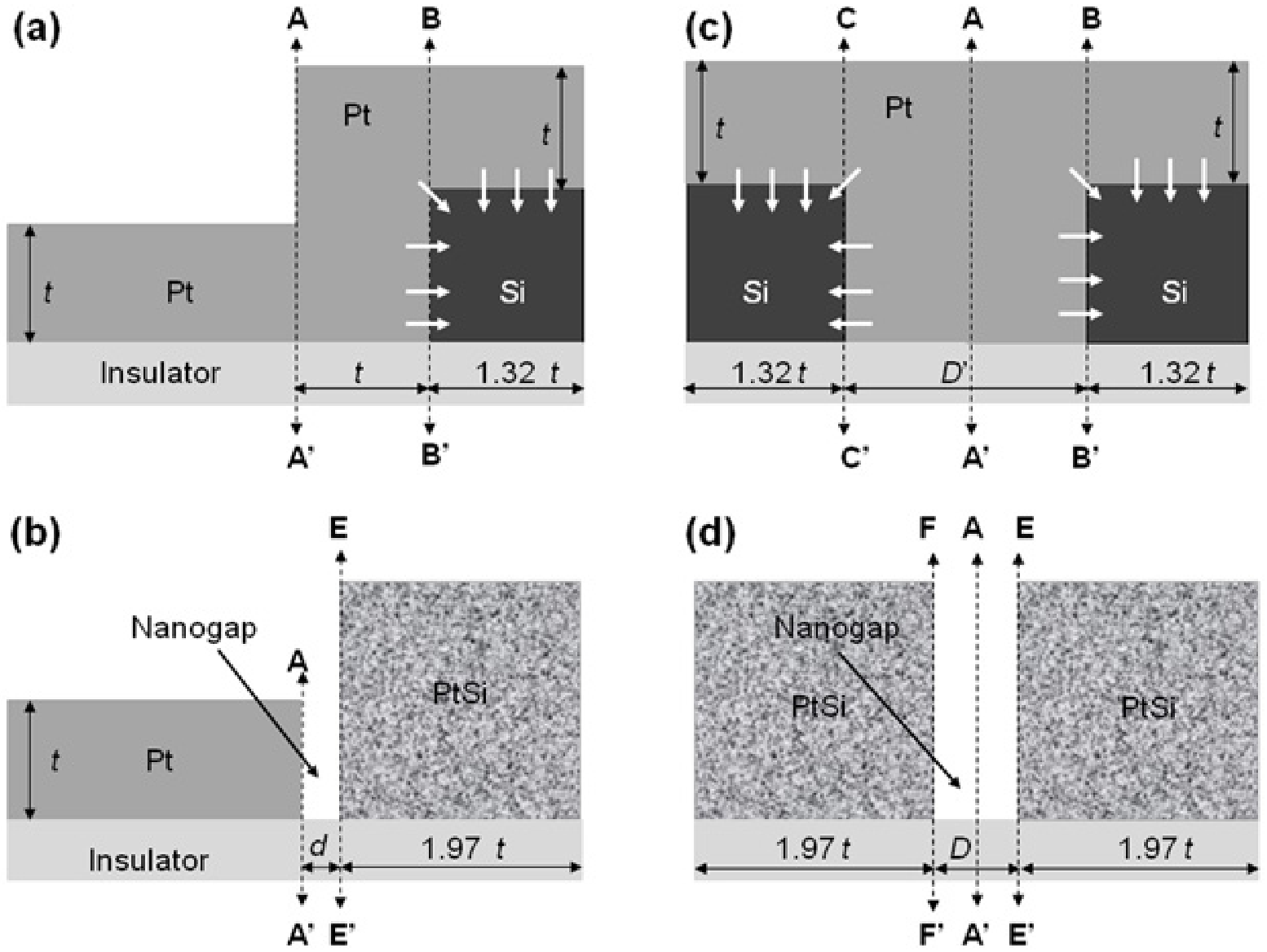}
\caption{}
\end{center}
\end{figure}

\newpage
\begin{figure}
\begin{center}
\includegraphics[width=8cm]{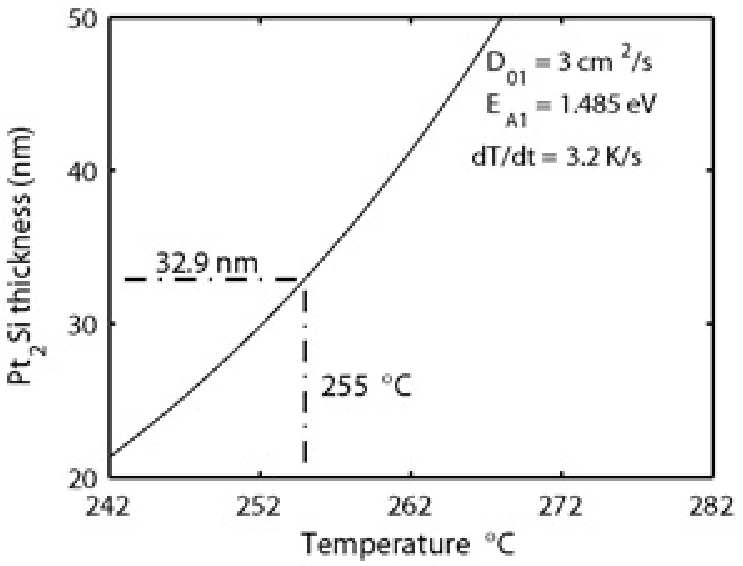}
\caption{}
\end{center}
\end{figure}

\newpage
\begin{figure}
\begin{center}
\includegraphics[width=16cm]{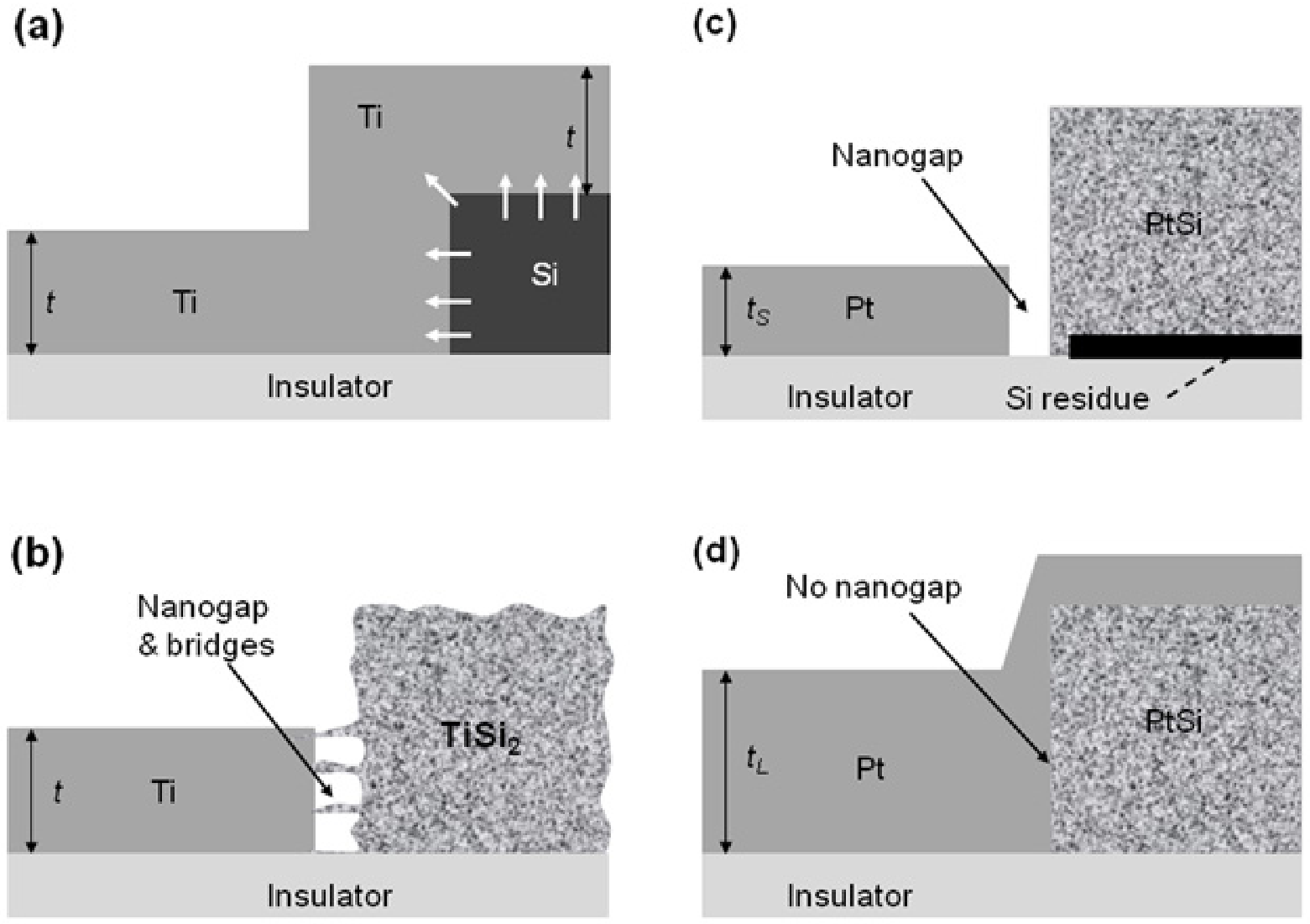}
\caption{}
\end{center}
\end{figure}

\newpage
\begin{figure}
\begin{center}
\includegraphics[width=16cm]{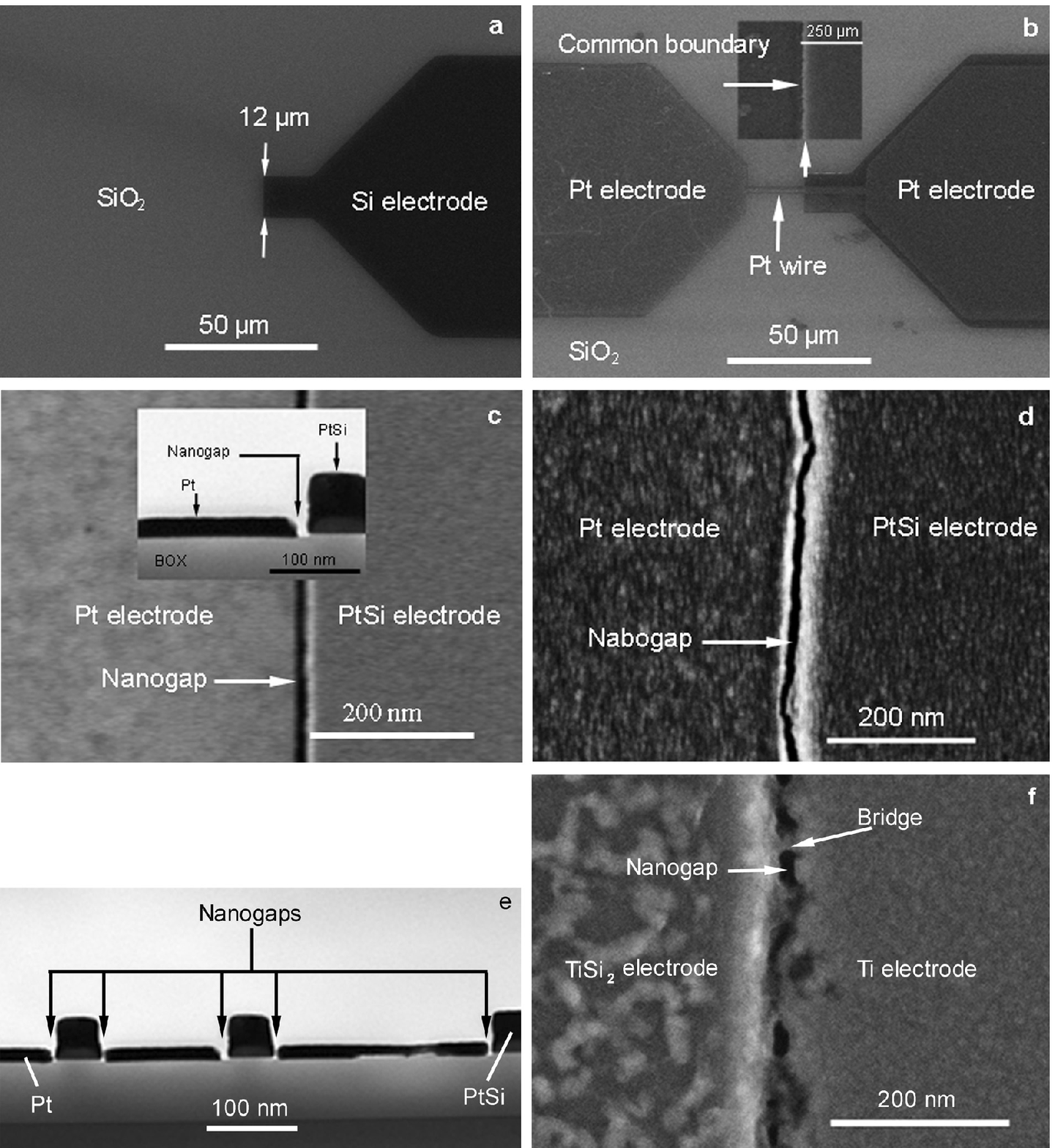}
\caption{}
\end{center}
\end{figure}

\newpage
\begin{figure}
\begin{center}
\includegraphics[width=16cm]{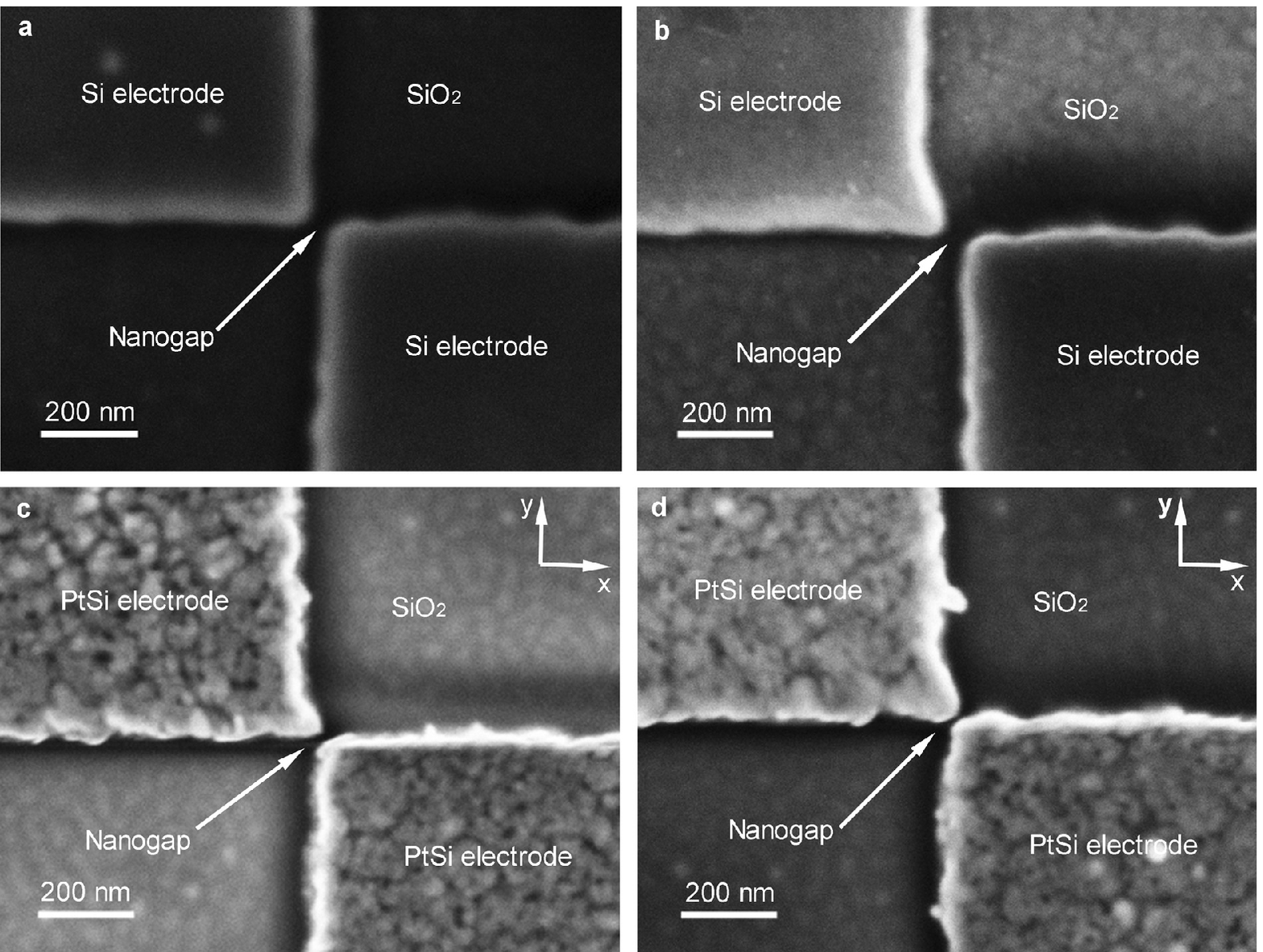}
\caption{}
\end{center}
\end{figure}

\newpage
\begin{figure}
\begin{center}
\includegraphics[width=8cm]{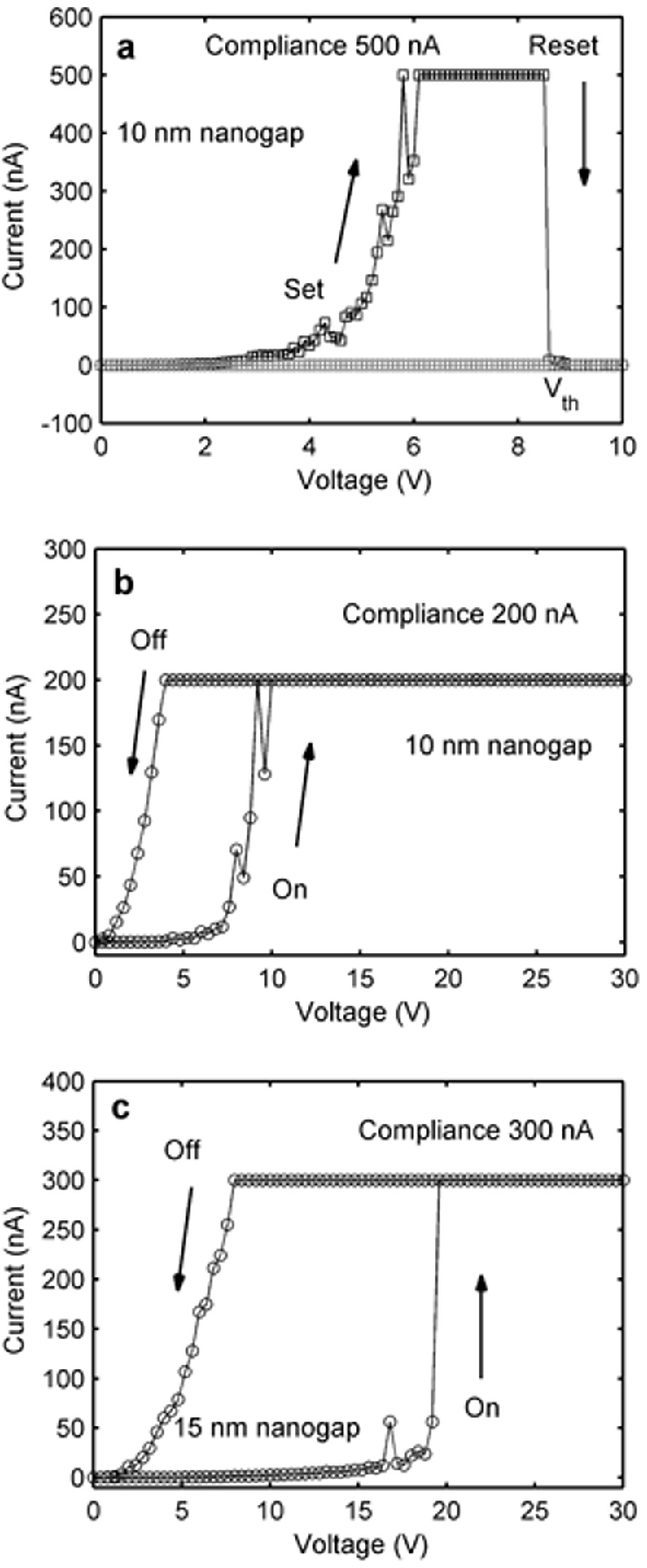}
\caption{}
\end{center}
\end{figure}

\end{document}